\documentstyle[11pt,a4,cite]{article}
\def\ga{\alpha}
\def\gb{\beta}

\def\ge{\epsilon}
\def\gg{\gamma}
\def\gd{\delta}

\def\gm{\mu}

\def\gp{\pi}

\def\gs{\sigma}
\def\gS{\Sigma}

\def\gL{\Lambda}

\def\delp{\partial_+}
\def\delm{\partial_-}
\def\part{\partial}
\def\delmu{\part_\gm}

\def\partk{\part_k}
\def\delr{\stackrel {\leftrightarrow}{\partial_\gm}}

\def\hlf{\frac{1}{2}}
\def\A0{A^{+}_0}
\def\psip{\psi_+}
\def\psin{\psi_-}
\def\psib{\bar{\psi}}
\def\psid{\psi^{\dag}}
\def\psipd{\psi_+^{\dagger}}
\def\psind{\psi_-^{\dagger}}

\def\xmin{x^{-}}
\def\ymin{y^{-}}
\def\xpl{x^+}
\def\xperp{x^{\perp}}

\newcommand{\nc}{\newcommand}
\nc{\uli}{\underline}
\def\ulix{\uli{x}}
\def\uliy{\uli{y}}

\def\ulip{\uli{p}}
\def\ulipp{\uli{p}^\prime}
\nc{\intgl}{\int\limits_{-L}^{+L}\!{dx^-\over{2L}}}
\nc{\intgp}{\int\limits_{-L_\bot}^{+L_\bot}\!d^2 x_\bot}
\nc{\intgly}{\int\limits_{-L}^{+L}\!{dy^-\over{2}}}
\nc{\intgv}{\int_V \!d^3 \ulix}
\nc{\zmint}{\int\limits_{-L}^{+L}\!{{dx^-}\over{\!2L}}}
\def\beq{\begin{equation}}
\def\eeq{\end{equation}}
\def\bea{\begin{eqnarray}}
\def\eea{\end{eqnarray}}
\begin{document}
\title{Fermionic Zero Modes and Spontaneous Symmetry Breaking on the 
Light Front }  
\author{\sl{L$\!\!$'ubom\'{\i}r Martinovi\v c} \\
Institute of Physics, Slovak Academy of Sciences \\
D\'ubravsk\'a cesta 9, 842 28 Bratislava, Slovakia 
\thanks{permanent address}\\
and\\
International Institute of Theoretical and Applied Physics\\
\medskip
Iowa State University, Ames, Iowa 50011, USA \\
and\\
{\sl James P. Vary} \\
Department of Physics and Astronomy \\
and\\
International Institute of Theoretical and Applied Physics, \\
Iowa State University, Ames, Iowa 50011, USA}  
\maketitle
\begin{abstract}

Spontaneous symmetry breaking is studied within a simple version of the 
light-front $O(2)$ sigma model with fermions. 
Its vacuum structure is derived by an implementation of global symmetries in 
terms of unitary operators in a finite volume with periodic fermi field. 
Due to the dynamical fermion zero mode, the vector and axial $U(1)$ 
charges do not annihilate the light-front vacuum. The latter is transformed 
into a continuous set of degenerate vacuum states, leading to the spontaneous 
breakdown of the axial symmetry. The existence of associated  massless 
Nambu-Goldstone boson is demonstrated.  
\end{abstract}
 

The phenomenon of spontaneous symmetry breaking represents a challenge in the 
light-front (LF) formulation of quantum field theory. In contrast to the usual 
quantization on space-like surfaces, the vacuum of the theory quantized on the 
surface of the constant LF time $x^+$ (i.e. on the light front) can be defined 
kinematically as a state with minimum (zero) longitudinal LF momentum $p^+$, 
since the operator $P^+$ has a positive spectrum \cite {LKS}. Thus, neglecting 
modes of quantum fields with $p^+ = 0$ (zero modes -- ZM), the vacuum state of 
even the interacting theory does 
not contain dynamical quanta. This ``triviality'' of the ground state is very 
advantageous for the Fock-state description of the bound states \cite{PB}, but 
it seems to forbid such important non-perturbative aspects  
like vacuum degeneracy and formation of condensates. Since there is no a priori
reason to expect any inconsistency \cite{LKS,Rohr} in Dirac's front  
form of relativistic dynamics \cite{Dir,Dir82}, it should be a sensible
strategy to look for a genuine LF description of symmetry breaking 
\cite{Yam} and related aspects of the vacuum structure \cite{LM1}, 
which would  
complement the usual space-like formulation based on a very complex 
dynamical vacuum state. 
Note in this context that, on the LF, in contrast to the space-like 
quantization \cite{Col,FPic}, even those charges which correspond to 
non-conserved currents do annihilate the vacuum in the continuum theory 
\cite{JS,Leutw}. Thus, one may expect similar ``surprises'' in other aspects 
of the LF field theory. 

A convenient regularized framework for studying these and related problems of 
non-perturbative nature is quantization in a finite volume with fields obeying 
periodic boundary conditions. It allows one to separate infrared aspects (ZM 
operators relevant for vacuum properties) from the remainder of the dynamics 
\cite{MY}. Note that to have a well-defined theory, one has to specify boundary conditions also in the continuum formulation \cite{Steinh,Heinz}.  
     
For self-interacting LF scalar theories a bosonic ZM is not a dynamical 
degree of freedom \cite{MY} but a constrained variable. Thus the vacuum 
remains indeed ``empty'' and one expects that physics of spontaneous symmetry 
breaking (SSB) is contained in solutions of a complicated 
operator ZM constraint \cite{Pin,Rg}. If a continuous symmetry is 
spontaneously broken, a massless Nambu-Goldstone (NG) boson should be present  
in the spectrum of states. However, as has been emphasized by Yamawaki and   
collaborators \cite{Yam}, the Goldstone theorem cannot exist on the 
light front as long as all charges annihilate the LF vacuum. Instead, 
a singular behaviour of the NG field and the charge non-conservation in the 
massless limit of a regularized theory has been identified as the manifestation 
of the NG phase in the LF scalar theories. 

The situation is different however for LF fermions in (3+1) dimensions. A  
massless fermion field, when quantized in a finite volume with periodic 
boundary conditions, contains a global ZM which is a dynamical variable.  
Recently, it has been demonstrated within the massive LF Schwinger 
model with antiperiodic fermion field that the residual symmetry under large 
gauge transformations, when realized quantum mechanically, gives rise to a 
non-trivial vacuum structure in terms of gauge-field zero mode as well as of 
fermion excitations \cite{LM1}. It is a purpose of the present work to 
demonstrate that dynamical fermion ZM provides a similar mechanism for a   
simple non-gauge field theory with fermions. Charges, which are the generators 
of global symmetries of the given system, contain a ZM part and 
consequently transform the trivial vacuum into a continuous set of 
degenerate vacuum states. This leads  
to a SSB in the usual sense \cite{NJL,Gold,GSW,Swieca,GHK,IZ,Strocchi}  
with non-zero vacuum expectation values of certain operators and a massless 
NG state in the spectrum of states. Much of what we demonstrate is of a 
rather general nature.  

In the LF field theory, the dynamical symmetry breaking \cite{NJL,Gold} has 
been studied so far within the usual mean-field approximation 
\cite{Heinzl,Itak} and  
also by means of Schwinger-Dyson equations \cite{Elkh}.  
 
To simplify our discussion of SSB in the LF field 
theory as much as possible, we will consider a version of the $O(2)$-symmetric 
sigma model with fermions \cite{Yam,IM} specified by the Lagrangian density 
\bea
{\cal L} & = & \psib\big({i \over 2}\gg^{\gm}\delr - m\big)\psi + 
\hlf(\part_\gm 
\gs \part^\gm \gs + \part_\gm \gp \part^\gm \gp) \nonumber \\
& - & \hlf \gm^2(\gs^2 + 
\gp^2) - g\psib(\gs + i\gg^5 \gp)\psi, 
\label{lagr}
\eea 
where the quartic self-interaction term for the scalar fields 
$\gs$ and $\gp$ has been omitted, because it is not relevant for our purpose.  
The Lagrangian (\ref{lagr}) is invariant under the global 
$U(1)$ transformation $\psi \rightarrow \exp(-i\ga)\psi$ and for $m=0$ also 
under the axial transformation
\bea
&&\psi \rightarrow \exp(-i\gb\gg^5)\psi,\;\;\psid \rightarrow \psid \exp(i\gb 
\gg^5),\label{axtrf}\\  
&&\gs \rightarrow \gs \cos 2\gb - \gp \sin 2\gb,\;\;\gp \rightarrow 
\gs \sin 2\gb + \gp \cos 2 \gb .
\label{axtrs}  
\eea
Rewriting the above Lagrangian in terms of the LF variables, one finds for 
the LF Hamiltonian 
\bea
P^- & = & \intgv\; \Big[(\partk \gs)^2 + (\partk \gp)^2 + \gm^2 (\gs^2 + \gp^2)
 + \psipd \big(m\gg^0 \nonumber \\
& - &i\ga^k\partk \big)\psin +  g\psipd\gg^0(\gs + i\gg^5\gp)\psin + 
h.c. \Big],  
\label{lfham}
\eea
where $d^3\ulix \equiv \hlf dx^- d^2\xperp$. 
Our convention for LF coordinates is $x^{\pm}=x^0 \pm x^3$,  
$p^\gm x_\gm = \hlf p^-\xpl+\ulip\ulix,\; \ulip\ulix=\hlf p^+ \xmin - 
x^\perp p^\perp, x^\perp p^\perp \equiv x^kp^k, k=1,2$ and $\xpl,p^-$ are 
the LF time and energy. Correspondingly, we define the Dirac matrices as 
$\gg^{\pm}=\gg^0 \pm \gg^3$, $\ga^k = \gg^0\gg^k$, the LF projection operators 
as $\gL_{\pm}=\hlf\gg^0\gg^{\pm}$ and $\gg^5 = i\gg^0\gg^1\gg^2\gg^3$. 
$\gL_{\pm}$ separate the fermi field into the independent component 
$\psip=\gL_+ \psi$ and the dependent one $\psin= \gL_-\psi$.
 
The infrared-regularized formulation is achieved by enclosing the system into 
a three-dimensional box $-L \leq \xmin \leq L,-L_\perp \leq x^k \leq L_\perp$
with volume $V=2L(2L_\perp)^2$ and by imposing  periodic boundary conditions 
for all fields in $\xmin,\xperp$. This leads to a decomposition of the fields 
into the zero-mode (subscript $0$) and normal-mode (NM, subscript n) parts. 
One finds that  
$\psin,\psind,\gs_0,\gp_0$ are non-dynamical fields with vanishing 
conjugate momenta, while $\psip,\psipd,\gs_n,\gp_n$ are dynamical. For a 
consistent quantization, one should apply the Dirac-Bergmann or a 
similar method suitable for systems with constraints. We will postpone that for 
a more detailed work \cite{ssbnext}, assuming here the standard 
(anti)commutators at $\xpl=0$:  
\bea
&&\{\psi_{+i}(\ulix),\psi^{\dagger}_{+j}(\uliy)\} = \hlf
\gd_{ij}\gd^3(\ulix - \uliy),\; i,j=1,4, 
\label{acr} \\
&&\big[\phi_n(\ulix),\delm \phi_n(\uliy)\big] =  
{i \over 2} \gd^3_n(\ulix-\uliy). 
\eea 
We are working in chiral representation with diagonal $\gg^5$ and $\phi = \gs$ 
or $\gp$. The anticommutator (\ref{acr}) can be derived 
by a direct calculation based on the field expansion  
\bea
&&\psip(\ulix) = \!\sum_{p^+,p^\perp \atop{s=\pm\hlf}} {u(s) \over 
\sqrt{V}}\big(b(\ulip,s)e^{-i\ulip\ulix} + d^{\dagger}(\ulip,-s)e^{
i\ulip\ulix}\big), \label{fexpansion} \\
&&\{b(\ulip,s),b^{\dagger}(\ulipp,s^\prime)\} = 
\{d(\ulip,s),d^{\dagger}(\ulipp,s^\prime)\} = \gd_{s,
s^\prime}\gd_{\ulip,\ulipp}.
\label{facr}
\eea
Here and in the Fourier representation of the periodic 
delta function $\gd^3(\ulix-\uliy)= \gd_0 + \gd_n^3(\ulix-\uliy), 
\gd_0={2 \over V}$, the summations run over discrete momenta 
$p^+=2\gp L^{-1}n,\;n=0,1\dots,N \rightarrow \infty,\;p^k =  
\gp L^{-1}_{\perp}n^k,\;n^k=0,\pm 1,\dots, \pm N_{\perp} \rightarrow \infty.$
The spinors are $u^{\dagger}(s=\hlf)=(1~0~0~0), 
u^{\dagger}(s=-\hlf)=(0~0~0~1)$ with $s$ being the LF helicity.  

The non-dynamical fields satisfy the constraints
\bea
&&2i\delm \psin=\left[m\gg^0 - i\ga^k\partk + g\gg^0\left(\gs + 
\gg^5\gp\right)\right]\psip, 
\label{psicon}\\
&& (\partk\partk - \gm^2)\gs_0  = g\intgl(\psipd\gg^0\psin + h.c.), 
\label{sicon}\\
&& (\partk\partk - \gm^2)\gp_0  = g\intgl(i \psipd \gg^0\gg^5\psin + h.c.).
\label{picon}
\eea 
The fermion constraint (\ref{psicon}) requires 
the dynamical fermion ZM $\psi_{+0}$ to vanish in the free massive theory. For 
the free massless fermi field, only the 
$\ulix$-independent {\it global} ZM is compatible with the constraint.
Decomposing $\gs_0,\gp_0$ into the {\it proper} zero modes \cite{AC2}  
$\gs_0(\xperp),\gp_0(\xperp)$ (which will not be needed here) and the global 
ZM $\hat{\gs}_0, \hat{\gp}_0$, the above constraints 
can be projected into three global-ZM sector relations.  
In Eqs.(\ref{sicon}) and (\ref{picon}), we have assumed the existence of the 
$\psi_{-0}$ zero mode, so the integrands are given by the diagonal 
combinations $\psi_{+0}\gg^0\psi_{-0} + \psi_{+n}\gg^0\psi_{-n}$, etc.   
Actually, by combining the global ZM constraints (with $m=0$) into one 
\bea
&&\Big(\psi^{\dagger}_{+0}
\gg^0\psi_{-0} + h.c.\Big)\psi_{+0} - \Big(\psi^{\dagger}_{+0}\gg^0\gg^5 
\psi_{-0} - h.c. \Big)\gg^5\psi_{+0} \nonumber \\ 
&& +\int_V{{d^3\ulix}\over V}\Big[\big(\psi^{\dagger}_ {+n}\gg^0 \psi_{-n}  +  
h.c.\big) - \big(\psi^{\dagger}_{+n}\gg^0\gg^5\psi_{-n}   
\nonumber \\ 
&& - h.c. \big)\gg^5 \Big]\psi_{+0}  =  
{\gm^2 \over g}\int_V{{d^3\ulix}\over V}\Big(\gs_n + i \gp_n\gg^5 \Big)
\psi_{+n},  
\label{combcon}
\eea
we see that non-zero $\psi_{-0}$ is {\it required} for consistency:   
setting $\psi_{-0}=0$ in Eq.(\ref{combcon}) yields an operator relation 
among independent fields which cannot be satisfied.  

$\psi_{-n}$ can be determined from the constraint (\ref{psicon}): 
\bea
&&\psi_{-n}(\ulix)\! =\! {1 \over{4i}}\intgly\ge_n(\xmin \!-\ymin)
\Big\{\big(m\gg^0 
-i\ga^k\partk\big)\psi_{+n}(\uliy) \nonumber \\ 
&&+ g\gg^0\Big[\left(\gs_n(\uliy) + i\gp_n(\uliy)\gg^5
\right) +  \left(\gs_0 + i\gp_0\gg^5 \right) \Big]\Big\} \psip(\uliy),\! 
\label{psinsol}
\eea
where $\ge_n(\xmin-\ymin)$ is the normal-mode part of the periodic sign 
function and $\uliy \equiv (\ymin,\xperp)$. Due to the presence of 
$\gs_0,\gp_0$, which in turn are 
given by their own constraints (\ref{sicon}),(\ref{picon}) depending on $\psi_
{-n}$, it is difficult to solve (\ref{psinsol}) in a closed form. Iterative 
solutions are possible and the lowest order one is obtained by setting 
$\gs_0=\gp_0=0$.  
 
While the free massive fermion Hamiltonian 
is, unlike the space-like quantization, symmetric under the axial vector  
transformations (\ref{axtrlf}) below \cite{Must}, the mass 
term in the $\psin$-constraint generates interaction terms  
which are proportional to $mg$ and which, due to an extra $\gg^0$, violate the 
axial symmetry explicitly. This is the reason why we shall set $m=0$ henceforth. 
Note however that the scalar fields have to be massive \cite{Yam} to avoid 
infrared problems.   

Inserting the $\psi_{-n}$-constraint into (\ref{lfham}), we find  
\bea 
&&P^-_{int}=\intgv\; \Big[ \gm^2(\gs_0^2 + \gp_0^2) + (\partk\gs_0)^2 + 
(\partk\gp_0)^2\Big] \nonumber \\
&&+ig \intgv\; \psipd(\ulix)\gS^{\dagger}(\ulix)  
\intgly \hlf\ge_n(\xmin \!\!-\ymin) \times \nonumber \\ 
&&\Big[ i\gg^k\partk\psi_{+n}(\ymin\!\!,\xperp) + h.c.   
-g\gS(\ymin\!\!,\xperp)\psip(\ymin\!\!,\xperp)\Big], 
\label{Hmassless}
\eea
where $\gS(\ulix) \equiv \gs(\ulix) + i\gp(\ulix)\gg^5$. 
It is not a closed expression due to the presence of $\hat{\gs}_0,\hat{\gp}_0, 
\gs_0(\xperp),\gp_0(\xperp)$. However, this is not an obstacle for
determining  the symmetry properties of the Hamiltonian, which are of 
primary importance in the present approach. First, we observe that the LF 
analogue of the axial vector transformation (\ref{axtrf}) is  
\beq  
\psip(\ulix) \rightarrow \exp{(-i\gb\gg^5)}\psip(\ulix),  
\label{axtrlf}
\eeq
while the NM fields $\gs_n,\gp_n$ transform according to (\ref{axtrs}). As for  
the constrained variables, we shall demand that $\psi_{-n}$ has a well defined 
transformation law, which is unambiguously fixed by the terms with $\ga^k, 
\gs_n$ and $\gp_n$ in the solution (\ref{psinsol}). It follows that  
$\gs_0 + i\gp_0\gg^5$ will transform exactly as $\gs_n + i \gp_n\gg^5$ and 
that the whole $\psi_{-n}$ will transform for $m=0$ in the same way as 
$\psip$. As a result, we find that $P^-_{int}$   
is invariant under $U_A(1)$ transformations in addition to $U(1)$.

These symmetries give rise to the conserved (normal-ordered) vector current 
$j^\gm = 
:\psid\gg^0\gg^\gm\psi:,~\delmu j^\mu = 0$ and the conserved axial-vector 
current $j_5^\gm = \psid\gg^0 \gg^\gm\gg^5\psi + 2 (\gs\part^\mu 
\gp - \gp\part^\mu \gs \big)$ ($\gm=+,-,k$): 
\beq
\delmu j^\mu_5 = 2m\Big(i\psipd \gg^0\gg^5\psin + h.c.\Big) = 0\;\; 
{\rm for}\;\; m=0.    
\eeq

They are implemented by the unitary operators  
$U(\ga)=\exp(-i\ga Q)$, $V(\gb)=\exp(-i\gb Q^5)$:
\bea
\psip(\ulix) \rightarrow e^{-i\ga}\psip(\ulix) = U(\ga)\psip(\ulix)U^{\dagger}
(\ga),\nonumber \\
\psip(\ulix)\rightarrow e^{-i\gb\gg^5}\psip(\ulix) = V(\gb)\psip(\ulix)
V^{\dagger}(\gb).
\label{u1}
\eea
While the NM parts of the charge operators $Q$ and $Q^5$ 
\bea 
Q = \intgv j^+(\ulix) = 2\intgv \psipd\psip,\\ 
Q^5 = 2\intgv \Big[\psipd\gg^5\psip + 2\Big(\gs_n\delm
\gp_n - \gp_n\delm\gs_n\Big)\Big]
\label{charges}
\eea  
are diagonal in creation and annihilation operators, the ZM parts,   
which do not vanish in the free nor the interacting  
theory, contain also off-diagonal terms   
\bea
Q_0 &=& \sum_{s} \big[b_0^{\dagger}(s)b_0(s) 
- d_0^{\dagger}(s)d_0(s) \nonumber \\ 
&+& b_0^{\dagger}(s)
d_0^{\dagger}(-s) + d_0(s)b_0(-s)\big],
\eea
\bea
Q_0^5 &=& \sum_{s}2s \big[b_0^{\dagger}(s)
b_0(s)
+ d_0^{\dagger}(s)d_0(s) \nonumber \\ 
&+& b_0^{\dagger}(s)
d_0^{\dagger}(-s) - d_0(s)b_0(-s)\big].
\eea  
The commuting ZM charges $Q_0, Q^5_0$ {\it do not} annihilate the LF 
vacuum $\vert 0 \rangle$ defined by $b(\ulip,s)\vert 0\rangle = d(\ulip,s)
\vert 0 \rangle = 0$. However, their vacuum expectation values  
are zero as they have to be. 
In this way, the vacuum of the model transforms under $U(\ga), 
V(\gb)$ as $\vert 0 \rangle \rightarrow \vert \ga \rangle = \exp(-i\ga Q_0) 
\vert 0 \rangle,~ \vert 0 \rangle \rightarrow \vert \gb \rangle  = \exp(-i\gb 
Q_0^5)\vert 0 \rangle $, where  
\bea
\vert \ga \rangle & = &\exp\left(-i\ga\sum_{s}\left[b_0^{\dagger}(s)
d_0^{\dagger}(-s) +h.c. \right]\right)\vert 0 \rangle, \\
\vert \gb \rangle & = & \exp\left(-i\gb\sum_{s}2s\left[b_0^{\dagger}(s)
d_0^{\dagger}(-s) +h.c. \right]\right)\vert 0 \rangle.
\label{degvac}
\eea
The vacua contain ZM fermion-antifermion pairs with opposite  
helicities. Due to Fermi-Dirac statistics, the number of such "Cooper pairs" 
cannot exceed two.  
 
Thus, the global symmetry of the Hamiltonian (\ref{Hmassless}) 
leads to an infinite set of translationally invariant 
states $\vert \ga,\gb\rangle = U(\ga)V(\gb)\vert 0 \rangle$ 
($P^+\vert \ga,\gb \rangle = P^\perp \vert 
\ga,\gb \rangle = 0$), labeled by two real parameters. 
Since $U(\ga), V(\gb)$ commute with $P^-$, the vacua are degenerate 
in the LF energy. The Fock space can be built from any of them since 
they are unitarily equivalent. 

We are in a position now to demonstrate the existence of the Goldstone theorem 
in the considered model. We have all the ingredients for the  
usual proof of the theorem \cite{GSW,Swieca,GHK,IZ,Strocchi}: the 
existence of the conserved current 
$j^\gm_5$, the operators $A$, namely  $\psib \psi = \psipd \gg^0 \psin + 
\psind \gg^0 \psip$ and $\psib\gg^5 \psi = \psipd \gg^0 \gg^5 \psin + \psind 
\gg^0 \gg^5 \psip$, which are non-invariant under the axial transformation 
\beq
A \rightarrow V(\gb)AV^{\dagger}(\gb) \neq A\Rightarrow \gd A = 
-i\gb[Q^5,A] \neq 0, 
\label{orderp}
\eeq
and the property $Q^5 \vert \ga,\gb \rangle = Q_0^5 \vert \ga,\gb 
\rangle \neq 0$. Of course, the above fermi bilinears are symmetric under 
$U(1)$, so the commutator $[Q,A]$ vanishes and there is no symmetry breaking 
associated with this symmetry. 

In a little more detail, from the axial current conservation 
and the periodicity in $\xmin,\xperp$ we get  
\beq
\delp \langle vac \vert \big[Q^5(\xpl),A \big]\vert vac \rangle = 0,\;\;
\vert vac \rangle \equiv \vert \ga,\gb \rangle
\label{ssbcom}
\eeq
in addition to
\beq
\langle vac \vert \big[Q^5(\xpl),A \big] \vert vac \rangle \neq 0. 
\label{ssbcon}
\eeq
These expressions imply that the the vacuum expectation value of the above 
commutator is a time-independent quantity. Note that the relation 
(\ref{ssbcon}) is only possible due to the fact that $Q^5$ does not annihilate 
the vacuum and this crucially depends on the existence of the ZM part of $Q^5$.
Inserting now a complete set of four-momentum eigenstates into the 
Eqs.(\ref{ssbcom}) and (\ref{ssbcon}) and using the translational invariance  
\beq 
e^{-iP_\gm x^\gm}\vert vac \rangle = \vert vac \rangle,\; 
j^+_5(x) = e^{-iP_\gm x^\gm} j^+_5(0) e^{iP_\gm x^\gm}
\label{translop}
\eeq 
we arrive in the usual way \cite{GSW,Swieca,IZ,Strocchi} to the conclusion 
that there must exist a state $\vert n \rangle = \vert G \rangle$ such, that 
\beq
\langle vac \vert A \vert G \rangle \langle G \vert j^+_5(0) \vert vac 
\rangle \neq 0 
\label{goldstone}
\eeq
with $P^-_{\mbox{\tiny{G}}} = 0\;{\rm for}\; 
P^+_{\mbox{\tiny{G}}}=P^\perp_{\mbox{\tiny{G}}} = 0.$ Thus,    
$M^2_{\mbox{\tiny{G}}} = P^+_{\mbox{\tiny{G}}}P^-_{\mbox{\tiny{G}}} - 
(P^\perp_{\mbox{\tiny{G}}})^2 = 0$.
From the infinitesimal rotation of the Fock vacuum we have explicitly  
\beq
Q_0^5 \vert 0 \rangle = \sum_{s} 2s b_0^{\dagger}(s)
d_0^{\dagger}(-s)\vert    
0 \rangle \equiv \vert G \rangle .
\label{nannih}
\eeq

Using the transformation law of the $\psi_\pm$ fields and the 
anticommutator (\ref{acr}), one can show that
the relation (\ref{ssbcon}) implies non-zero vacuum expectation values of  
the operators $A$ \cite{ssbnext}. 
They will depend on the coupling constant through 
$\psi_{-n}$. To obtain quantitative results, one has to solve approximately the 
constraint (\ref{psicon}) \cite{IM}.

To summarize, we have demonstrated that spontaneous symmetry breaking
can occur  in the {\it finite-volume} formulation of the fermionic LF field
theory. While in contrast with the usual expectation within the
space-like field  theory (see \cite{Mir}, e.g.), this is related to the 
explicit presence of a dynamical fermion zero mode in the finite-volume 
LF quantization. One of 
the advantages of this infrared-regularized formulation is that one does not 
need to introduce test functions and complicated definitions of  
operators to obtain a mathematically rigorous framework \cite{Swieca}. For 
example, contrary to the standard infinite-volume formulation, the norm of 
the state $Q^5\vert vac \rangle = Q_0^5 \vert vac \rangle$ is finite and 
{\it volume-independent}. However, the issue of continuum limit 
and volume independence of the physical picture obtained 
in a finite volume requires a further study.  

In the usual treatment of fermionic theories \cite{NJL,Mir,HKuni}, the 
considered vacua, related by a canonical transformation, are the free-field 
vacua corresponding to fermion fields with different masses. In the LF
picture, such vacua are unitarily equivalent \cite{LKS}. 
Our approach relates the vacuum 
degeneracy to the unitary operators implementing the symmetries, making use of 
the ``triviality'' of the LF vacuum in the sector of normal Fourier modes.  

Nevertheless, there are still a few aspects of the present approach  
that have to be understood better. First, one has to perform a full 
constrained quantization of the model to derive the (anti)commutation 
relations for all relevant (ZM) degrees of freedom. Also, the connection of 
our picture with the standard one, based on the mean-field approximation and 
the new vacuum with lower energy above the critical coupling, has to be 
clarified. 


This work has been supported by the grants VEGA 2/5085/98,  
NSF No. INT-9515511 and by the U.S. 
Department of Energy, Grant No. DE-FG02-87ER40371.

\end{document}